\documentclass[12pt, prd, reprint, twocolumns, nofootinbib]{revtex4-1}
\usepackage{lineno,hyperref}
\usepackage{amsmath}
\usepackage{color,ulem, graphicx}
\usepackage{cancel}
\usepackage{ amssymb }

\begin{document}

\title{R-parity Violating Supersymmetric Explanation of the Anomalous Events at ANITA}

%% Group authors per affiliation:
\author{Jack H. Collins$^{1,2}$}
\author{P. S. Bhupal Dev$^3$}  
\author{Yicong Sui$^3$}

\address{$^1$Maryland Center for Fundamental Physics, Department of Physics, University of Maryland, College Park, MD 20742, USA}
\address{$^2$Department of Physics and Astronomy, Johns Hopkins University, Baltimore, MD 21218, USA}
\address{$^3$Department of Physics and McDonnell Center for the Space Sciences, Washington University, St. Louis, MO 63130, USA}
\begin{abstract}
The ANITA balloon experiment has observed two EeV-energy, upgoing events originating from well below the horizon. This is puzzling, because (i) no Standard Model (SM) particle is expected to survive passage through Earth at such energies and incident angles, and (ii) no such events were reported by IceCube. In this paper, we address both these issues by invoking a beyond SM interpretation of the EeV events as due to the decay of a long-lived bino in the R-parity violating (RPV) supersymmetry. In particular, a TeV-scale slepton/squark can be resonantly produced through the interaction of the EeV neutrino with electrons/nucleons inside Earth, that decays to a light, long-lived bino, which survives the propagation through Earth matter before decaying back to neutrinos, leptons and/or quarks, thus producing upgoing air showers in the atmosphere near ANITA. We find that the ANITA events can be explained with a GeV-scale bino and ${\cal O}(0.1)$ RPV couplings, which are consistent with all existing high and low-energy constraints. We also find that an isotropic neutrino flux is inadequate for { a beyond the SM explanation of this kind}, and an anisotropic flux must be invoked. Finally, we also address the apparent tension of these observations with IceCube. Various aspects of our interpretation are testable in the near future at different frontiers, such as by the LHC, Belle II, ANITA-IV and IceCube. 
\end{abstract}

%\begin{keyword}
%\texttt{elsarticle.cls}\sep \LaTeX\sep Elsevier \sep template
%\MSC[2010] 00-01\sep  99-00
%\end{keyword}

%\end{frontmatter}

\maketitle

\section{Introduction}
The Antarctic Impulsive Transient Antenna (ANITA) collaboration has recently reported two  anomalous upward-going ultra-high energy cosmic ray (UHECR) air shower events with deposited shower energies of $0.6\pm 0.4$ EeV and $0.56^{+0.3}_{-0.2}$ EeV, respectively~\cite{Gorham:2016zah, Gorham:2018ydl}. Both events, one from ANITA-I~\cite{Gorham:2016zah} and another from ANITA-III~\cite{Gorham:2018ydl}, %\footnote{The trigger algorithm used for ANITA-II was not sensitive to these events~\cite{Gorham:2016zah}.} 
originate from well below the horizon, with elevation angles of $(-27.4\pm 0.3)^\circ$ and $(-35.0\pm 0.3)^\circ$, respectively. They do not exhibit phase inversion due to Earth's geomagnetic effects -- a primary characteristic of conventional downgoing UHECR air showers which produce downgoing radio impulses that are reflected off the Antarctic ice surface.  Potential background events from anthropogenic radio signals that might mimic the UHECR characteristics, or unknown processes that might lead to a non-inverted polarity on reflection from the ice cap are estimated to be $\leq 0.015$. This leads to $\gtrsim 3\sigma$ evidence for the interpretation of the two anomalous events as due to direct upward-moving Earth-emergent UHECR-like air showers above the ice surface~\cite{Gorham:2018ydl}.

However, such an interpretation faces severe challenges within the known Standard Model (SM) framework, because no SM particle is expected to survive passage through Earth a chord distance of $\sim 7000$ km (corresponding to the observed zenith angles of the two events) at EeV energies. In particular, the interpretation of these events as $\tau$-lepton decay-induced air showers at or near the ice surface arising from a diffuse UHE flux of cosmic $\nu_\tau$ is strongly disfavored due to their mean interaction length of only $\sim 300$ km. Even including the effect of $\nu_\tau$ regeneration~\cite{Halzen:1998be,Bugaev:2003sw, Bigas:2008sw, Jeong:2017mzv}, the resulting survival probability over the chord length of the ANITA events with energy greater than $0.1 \; \text{EeV}$ is $<10^{-6}$~\cite{Fox:2018syq}, largely due to $\tau$-lepton energy loss inside Earth because of ionization, $e^+e^-$ pair production, bremsstrahlung, and photonuclear interactions~\cite{Alvarez-Muniz:2017mpk}, thereby excluding the SM interpretation at $5.8\sigma$ confidence.  A possible way out is by invoking significant suppression of the deep-inelastic neutrino-nucleon cross section above EeV~\cite{Connolly:2011vc, Chen:2013dza, Albacete:2015zra, Bustamante:2017xuy, Bertone:2018dse} due to gluon saturation at small Bjorken-$x <10^{-6}$~\cite{Henley:2005ms}. This will likely decrease the exponential attenuation of the Earth-crossing neutrino flux by at most a factor of 2-3~\cite{Armesto:2007tg, Illarionov:2011wc, Goncalves:2015fua}, whereas an order of magnitude or more suppression is needed to explain the two ANITA events. 

Another explanation of the anomalous events within the SM framework was proposed in terms of the transition radiation from a particle shower crossing the Earth-air interface and induced by an Earth-skimming neutrino~\cite{Motloch:2016yic}. In  this  model,  the  plane-of-polarization correlation to geomagnetic angles would be coincidental.  Since both ANITA events are well-correlated to the local geomagnetic angle, and are consistent within 3$^\circ$-5$^\circ$ of measurement error,  coincidental  alignment  for  both  is possible only  at  the  few \% level~\cite{Gorham:2018ydl}. Moreover, the diffuse neutrino flux necessary for this explanation to work is in tension with the current best limits from the Pierre Auger~\cite{Aab:2015kma}, IceCube~\cite{Aartsen:2018vtx} and ANITA~\cite{Allison:2018cxu} data.  

Several beyond the SM (BSM) interpretations of the ANITA anomalous events, such as sterile neutrino mixing~\cite{Cherry:2018rxj, Huang:2018als}, heavy dark matter~\cite{Anchordoqui:2018ucj, Dudas:2018npp} and stau~\cite{Connolly:2018ewv,Fox:2018syq} decays, have also been discussed. All of these explanations assume that the showers observed by ANITA were initiated by the hadronic decays of a $\tau$-lepton. However, a major challenge for any BSM interpretation in which the ANITA events are initiated by a decaying $\tau$ lepton is to explain the apparent discrepancy with the null observation of any comparably energetic and steeply inclined throughgoing track events at IceCube~\cite{Aartsen:2017mau}\footnote{Though it is worth noting that there are three IceCube events which could be interpreted as throughgoing $\tau$ tracks with energy 10 to 100 PeV and inclined at angles 10 to 30 degrees below the horizontal \cite{Fox:2018syq,Kistler:2016ask}.}, which has been operating at its design sensitivity for more than nine years, as compared to ANITA's { approximately two months exposure}. With an effective area of 1 km$^2$ (as compared to ANITA's 4 km$^2$) at EeV energies, the IceCube exposure is almost 12 times that of ANITA. Based on this argument, it was pointed out~\cite{Huang:2018als} that the sterile neutrino explanation~\cite{Cherry:2018rxj} is in strong tension with IceCube. The same conclusion holds for the hypothesis of quasi-stable dark mater decay inside the Earth~\cite{Anchordoqui:2018ucj}, and for the stau-based proposals~\cite{Connolly:2018ewv,Fox:2018syq}. Moreover, as pointed out in Ref.~\cite{Dudas:2018npp}, given the local dark matter density of 0.3~GeV \. cm$^{-3}$, the capture rate of an EeV-scale decaying dark matter is {very} low, corresponding to one dark matter decay every 137 years in the entire volume of the Earth.

%As for the stau (or any long-lived charged particle) explanation~\cite{Connolly:2018ewv,Fox:2018syq}, one has to properly account for the energy loss along its journey through the Earth matter due to electromagnetic, photonuclear and $e^+e^-$ pair production processes~\cite{Albuquerque:2003mi, Reno:2005si, Albuquerque:2006am, Huang:2006ie} (see also Refs.~\cite{Ahlers:2007js, Ando:2007ds, Canadas:2008ey} for some related discussion). Using the semi-analytic formula for the average range of stau inside Earth given in Ref.~\cite{Reno:2005si}, we estimate that a TeV-mass stau with EeV-energy can survive passage through rock up to about 3000 km before losing significant fraction of its energy.\footnote{Ref.~\cite{Connolly:2018ewv} quotes the range as $10^4$ km from Ref.~\cite{Reno:2005si}, but this is the water equivalent range and it will be roughly 3-4 times smaller for propagation through Earth, depending on whether the particle goes through Earth's core or mantle.} This estimated range just falls short of the $\sim 7000$ km chord length needed to explain the ANITA anomalous events. 

The assumption that the upward going showers observed by ANITA were initiated by $\tau$-leptons may be premature, since it is not clear how the ANITA experiment would distinguish between showers initiated by different kinds of particle decays on an event-by-event basis. The decays of a highly boosted BSM particle directly into hadrons, electrons, or photons would also result in the production of an impulsive radio cone, and this might give rise to miscalibrated energy measurement or effective area prediction when interpreted in terms of a $\tau$-hypothesis. Moreover, all the BSM scenarios discussed above assume an isotropic flux of incident neutrinos, which has serious problems producing the observed arrival directions at ANITA without overproducing at shallower angles. 
%As we confirm in this paper, an isotropic Greisen-Zatsepin-Kuzmin (GZK) neutrino flux cannot account for the ANITA anomalous events, irrespective of the underlying BSM model. 
As we discuss in this paper, it is difficult to account for the ANITA anomalous events using an isotropic Greisen-Zatsepin-Kuzmin (GZK) neutrino flux. 
%Taking the ANITA events at face value, one needs to consider a neutrino flux at least two orders of magnitude larger than the GZK flux, which is clearly ruled out by the constraints from Pierre Auger~\cite{Aab:2015kma}, IceCube~\cite{Aartsen:2018vtx} and ANITA~\cite{Allison:2018cxu}.
We therefore consider an anisotropic flux to fit the ANITA events and show that this is so far consistent with the existing searches for potential candidate transient sources in the northern sky.    

We then propose a new BSM solution to the ANITA puzzle in terms of a long-lived {\it neutral} particle. In particular, we advocate a GeV-scale bino in $R$-parity violating supersymmetry (RPV-SUSY) as a natural candidate for this purpose. Our solution has several advantages over the other BSM explanations entertained earlier: 
\begin{itemize}
\item [(i)] The RPV couplings allow the on-shell, resonant production of a TeV-scale squark/slepton from neutrino-nucleon/electron scattering inside Earth, thereby naturally enhancing the signal cross section. 

\item [(ii)] The squark/slepton decay inside the Earth can produce a pure bino which interacts with the SM fermions only via the $U(1)_Y$ gauge interactions and heavier supersymmetric particles, and therefore, can easily travel through thousands of km inside the Earth without significant energy loss. 
 %, unlike tau, stau or any other electrically-charged particle. 
 
\item [(iii)] For a suitable, yet realistic sparticle mass spectrum and couplings consistent with all existing low and high-energy constraints, we find some parameter space where the bino is sufficiently long-lived (with proper lifetime of order of ns) and decays to SM fermions at or near the exit-surface of Earth to induce the air shower observed by ANITA. 

\item [(iv)] For $LLE$-type RPV couplings, the bino decays to a $\tau$-lepton (if kinematically allowed) and electron, either of which could induce the air shower seen by ANITA, whereas for $LQD$-type couplings, the bino directly decays to quarks and neutrino, which mimic the SM $\tau$-decay. In the latter case, there exist some parameter space for which no throughgoing track events are predicted at IceCube.
\end{itemize}
 %In either case, we can explain the apparent absence of events at IceCube.    

 %An even more severe problem overlooked by most is whether GZK flux is indeed the reason for such anomalies since either AGN, GRB, blazars or GC could very well be possible point sources accounting for the events. 

%Our paper is organized as follows.  In Sec. \ref{s2} we first review the difficulties of interpreting the ANITA events in either SM or BSM way. Then we introduce our bino particle mechanism based on RPV SUSY which will successfully explain both the survival problem and the contradictions between ANITA and IceCube. After setting up the model, we proceeds in Sec. \ref{s3} to verify our model and find out the insufficiency in general for GZK to explain ANITA events. Then we provide possible alternative anisotropic sources interpretation and a favored region of the new physics parameters are given via a fitting to ANITA events. In Sec. \ref{s4}, we conclude that our model is by far the only model that reconcile most of the contradictions. \blue{To be finalized at the end.}

\section{The Model Setup}\label{sec:model}
%In order to satisfy all these conditions, we assume this $\chi$ to be a lightest sparticle(LSP) bino in RPV SUSY model\cite{Carena:1998gd} 
We start with the general trilinear RPV superpotential in the Minimal Supersymmetric SM~\cite{Barbier:2004ez}: 
%with general RPV terms:
\begin{equation}
W_{\cancel{R}} \ = \ \frac{1}{2}\lambda_{ijk} L_i L_j E_k^c + \lambda'_{ijk} L_i Q_j D_k^c+\frac{1}{2}\lambda''_{ijk} U_i^c D_j^c D_k^c \, ,
\label{eq:sup}
\end{equation}
where $L_i\ni (\nu_i,e_i)_L$ and $Q_i \ni (u_i,d_i)_L$ are the $SU(2)_L$-doublet and $U_i^c$,  $D_i^c$, $E_i^c$ are the $SU(2)_L$-singlet chiral superfields, and $i, j, k = 1,2,3$ are the generation indices. Here we have suppressed all gauge indices for brevity. Note that $SU(2)_L$ and $SU(3)_c$ gauge invariance enforce antisymmetry of the $\lambda_{ijk}$ and $\lambda''_{ijk}$ couplings with respect to their $i,j$ or $j,k$ indices, respectively. Since we are interested in the UHE neutrino interactions with matter, we will only consider the $\lambda$ and $\lambda'$-terms, one at a time.\footnote{In presence of the $\lambda'$-terms, the $\lambda''$-terms can be explicitly forbidden, e.g. by imposing baryon triality~\cite{Ibanez:1991hv}, in order to avoid dangerous proton decay operators~\cite{Smirnov:1996bg, Nath:2006ut}.}  

\subsection{$LLE$ Contribution}

\begin{figure}[t!]
	\centering
	\includegraphics[width=0.48\textwidth]{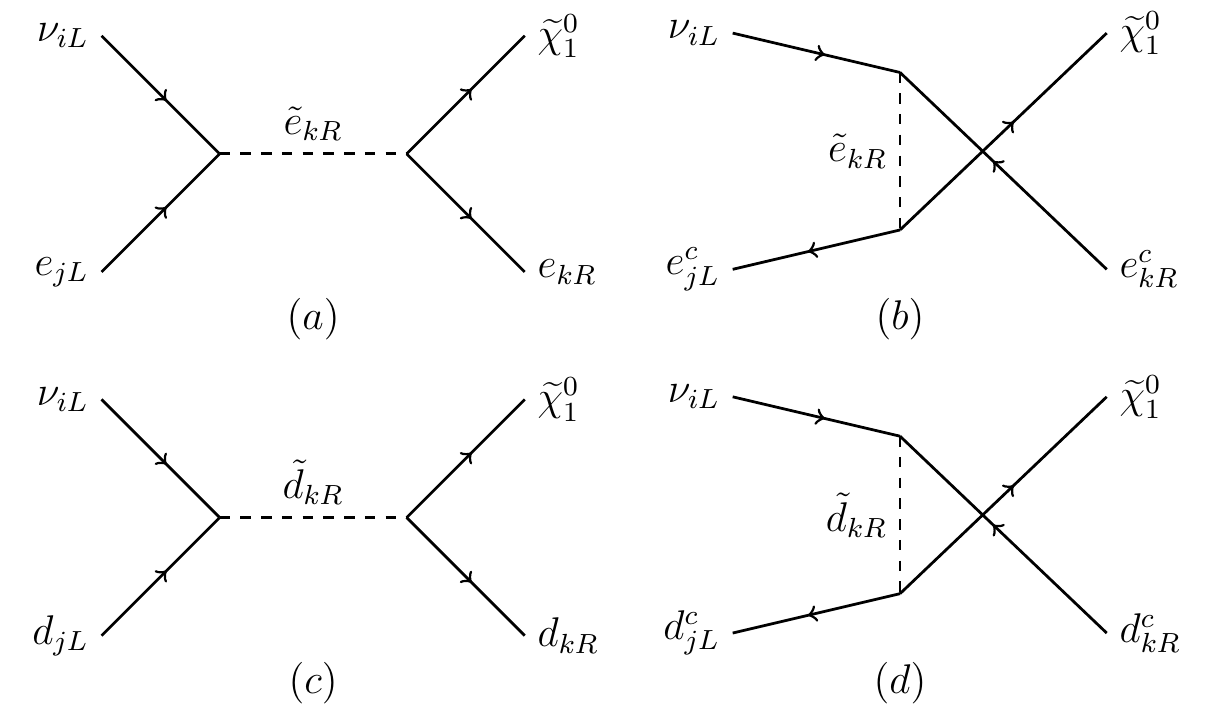}
	\caption{Representative Feynman diagrams for the neutrino-electron (top) and neutrino-nucleon (bottom) interactions via RH-sfermion mediation in our RPV-SUSY scenario to produce a long-lived bino. Similar diagrams exist for LH-sfermion mediation, which are not shown here, but included in the calculation. }
	\label{fig:feynman}
\end{figure}

Let us consider the $\lambda$-terms first. Expanding them in Eq.~\eqref{eq:sup}, we obtain the Lagrangian 
\begin{align}\label{eq:lag1}
\mathcal{L}_{LLE} \ = \ & \frac{1}{2}\lambda_{ijk}\bigg[ \widetilde{\nu}_{iL} \bar{e}_{kR} e_{jL} +\widetilde{e}_{iL} \bar{e}_{kR}\nu_{jL} +\widetilde{e}_{kR}^{*} \bar{\nu}_{iL}^c e_{jL} \nonumber \\
& \qquad \qquad - (i\leftrightarrow j) \bigg]+{\rm H.c.}
\end{align}
With these interactions, we can have new contributions to the (anti)neutrino-electron scattering inside Earth, as shown in Fig.~\ref{fig:feynman} (top panel). In particular, given enough energy of the incoming (anti)neutrino, this will lead to the resonant production of a left-handed (LH) slepton through the second term in Eq.~\eqref{eq:lag1}, and similarly a right-handed (RH) slepton through the third term in Eq.~\eqref{eq:lag1}. For an incoming neutrino energy $E_\nu$, the slepton mass at which the resonance occurs is simply $m_{\widetilde{e}_i}=\sqrt{s}=\sqrt{2E_\nu m_e}$~\cite{Carena:1998gd, Dev:2016uxj}, where $s$ is the center-of-mass energy. This is reminiscent of the Glashow resonance in the SM, where an on-shell $W$ boson is produced from the $\bar{\nu}_e-e$ scattering with an initial neutrino energy of $E_\nu=m_W^2/2m_e=6.3$ PeV~\cite{Glashow:1960zz}.  

Once produced, the slepton can decay back to an electron and neutrino through the same RPV interaction in Eq.~\eqref{eq:lag1} or to the corresponding lepton and neutralino through gauge interactions. { The slepton might in principle be from any generation, though here we will make the assumption that the slepton is a stau ($\widetilde{\tau}$)}, and also assume the lightest neutralino ($\chi_1^0$) to be much lighter than the stau, so that the decay $\widetilde{\tau}\to \tau \widetilde{\chi}_1^0$ is kinematically allowed. All other sparticles are assumed to be heavier than the stau and do not play any role in our analysis, except for the gravitino ($\widetilde{G}$), which could be the lightest supersymmetric particle (LSP) and plays the role of dark matter in this scenario.\footnote{Gravitino LSP and bino next-to-LSP (NLSP) can be realized, e.g. in natural gauge mediation without gaugino unification~\cite{Barnard:2012au}.}

The cross-section for $\nu e\to \widetilde{\tau}\to \tau\widetilde{\chi}_1^0$ production, which can be approximated by a Breit-Wigner formula close to the $s$-pole with $s\to m^2_{\widetilde{\tau}}$, is given by~\cite{Carena:1998gd} as:
%{\color{blue} \emph{(Not sure about this, check email)}}

%\begin{align}
%\sigma_{LLE} \ = \ \frac{8\pi s}{m^2_{\widetilde{\tau}}} \frac{\Gamma(\widetilde{\tau}\to e\nu)\ \Gamma(\widetilde{\tau}\to \tau\widetilde{\chi}_1^0)}{(s-m^2_{\widetilde{\tau}})^2+m_{\widetilde{\tau}}^2\Gamma_{\widetilde{\tau}}^2}\left(\frac{s-m^2_{\widetilde{\chi}_1^0}}{m^2_{\widetilde{\tau}}-m^2_{\widetilde{\chi}_1^0}}\right)^2 \, ,
%\end{align}
%which can be approximated by a Breit-Wigner formula close to the $s$-pole with $s\to m^2_{\widetilde{\tau}}$: 
\begin{align}
\sigma_{LLE}& \ \simeq \ \frac{8\pi}{m_{\widetilde{\tau}}^2}{\rm Br}(\widetilde{\tau}\to\nu e)\ {\rm Br}(\widetilde{\tau}\to\widetilde{\chi}_1^0\tau) \nonumber \\
& \ = \ \frac{8\pi}{m_{\widetilde{\tau}}^2}\frac{\lvert \lambda_{ijk}\rvert^2 g'^2}{(\lvert\lambda_{ijk}\rvert^2+g'^2)^2} \, ,\label{RPVXS}
\end{align} 
where $g'\equiv e/\cos\theta_w$ is the $U(1)_Y$ gauge coupling ($e$ being the electromagnetic coupling and $\theta_w$ the weak mixing angle), and $j=1$, $k=3$ or vice versa, depending on whether it is the RH or LH-slepton resonance, respectively. The index $i$ for the incoming neutrino is free and we will assume a democratic flux ratio $1:1:1$ for $\nu_e:\nu_\mu:\nu_\tau$ and similarly for antineutrinos, as expected for a typical astrophysical neutrino flux with $1:2:0$ flavor composition at the source~\cite{Learned:1994wg}. 

\begin{figure}[t!]
	\centering
	\includegraphics[width=0.48\textwidth]{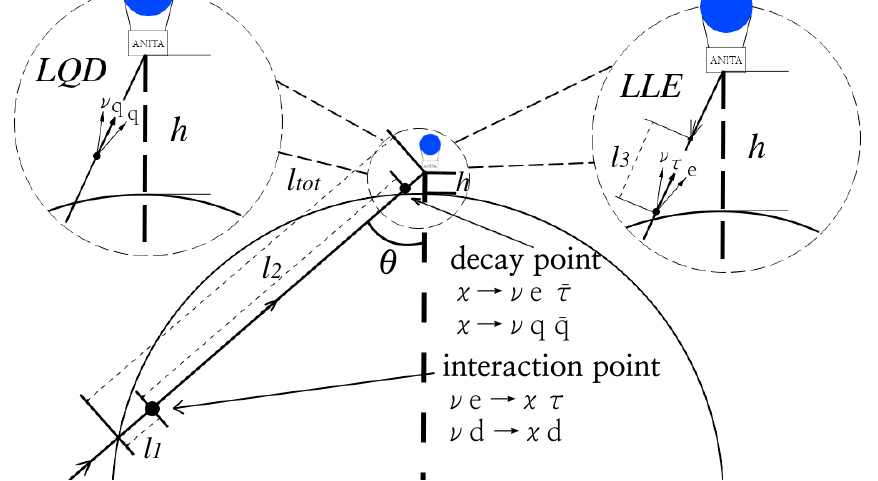}
	\caption{A sketch of our model setup. The incoming UHE neutrino interacts with electron or quark inside the Earth within a distance $l_1$ and produces a bino through the diagrams shown in Fig.~\ref{fig:feynman}. The bino travels a distance $l_2$, before decaying to $\nu e^- \tau^+$ or $\nu q\bar{q}$ close to the surface, which induces the air shower seen by ANITA.  Here, $l_{\rm tot}$ is the total distance between the point where the UHE neutrino enters Earth to the ANITA detector, located at a height $h$ above Earth's surface, and $\theta$ is the incoming angle of the neutrino with respect to the vertical direction.}
	\label{fig:sketch}
\end{figure}

We assume the bino is long-lived enough to survive its passage through Earth, before decaying close to or at the surface of exit. It can decay back to the $\tau$-lepton and an off-shell stau, leading to a 3-body final state: $\widetilde{\chi}_1^0\to \tau^+ \widetilde{\tau}^{*-} \to \tau^+ e^- \bar{\nu}$. In principle, the upgoing shower may be initiated either directly by the electron, or by the subsequent decay of the $\tau$, as shown schematically in Fig.~\ref{fig:sketch}. %Since ANITA has only presented the inferred energy and effective area for the $\tau$ decay hypothesis, we limit our subsequent analysis to the assumption that the events are initiated only by the $\tau$.
%It is this final-state $\tau$ that induces the upgoing air shower seen by ANITA in our model, as shown schematically in Fig.~\ref{fig:sketch}.
In the limit $m_{\widetilde{\chi}_1^0}\ll m_{\widetilde{\tau}}$, the 3-body decay rate can be estimated as 
\begin{equation}\label{decaywidth}
\Gamma(\widetilde{\chi}_1^0\to\tau^- e^+ \bar{\nu}) \ \simeq \ \frac{g'^2|\lambda_{ijk}|^2}{512\pi^3}\frac{m_{\widetilde{\chi}_1^0}^5}{m_{\widetilde{\tau}}^4} \, .
\end{equation}
%where $\alpha\equiv e^2/4\pi$ is the fine-structure constant. 

According to the geometry shown in Fig.~\ref{fig:sketch}, the incident neutrino travels a short distance $l_1$ inside Earth, and the remaining distance $l_2$ is traveled by the bino. In the limit $l_1\ll l_2$, we can approximate $l_2$ as the chord length of $\sim 7000$ km required for the two ANITA events, which translates into the mean lifetime of bino in the lab frame
%its rest frame
as $\tau_{\widetilde{\chi}_1^0}^{\rm lab}\sim 0.022$ s. From the decay kinematics of the event, we estimate that the incoming neutrino energy should be roughly four times the detected energy at ANITA. Given that the two ANITA events had an average energy of 0.5 EeV, the initial neutrino energy should be $E_\nu\sim$ 2 EeV. Then the resonance condition fixes the stau mass: $m_{\widetilde{\tau}}=\sqrt{2E_\nu m_e}\simeq 2$ TeV. Substituting this in Eq.~\eqref{decaywidth}, we find that for a typical value of $|\lambda|\sim 0.1$, as allowed by current experimental constraints~\cite{Kao:2009fg}, one needs a light bino of mass $m_{\widetilde{\chi}_1^0}\sim 8$ GeV. A more accurate calculation of the allowed range in the $(m_{\widetilde{\chi}_1^0}, |\lambda|)$ plane, taking into account all statistical interaction/decay probabilities for the neutrino, bino and tau, will be presented in a later section.   

We should mention here that for gravitino LSP and bino NLSP, the bino can also have a 2-body decay into a photon and a gravitino via its photino component and the decay rate is given by~\cite{Covi:2009bk}
\begin{align}\label{decaywidth2}
\Gamma(\widetilde{\chi}_1^0\to \widetilde{G}\gamma) \ = \ \frac{\cos^2\theta_w}{48\pi M_{\rm Pl}^2}\frac{m^3_{\widetilde{\chi}_1^0}}{x_{3/2}^2}\left(1-x^2_{3/2}\right)^3\left(1+3x^2_{3/2}\right) \, ,
\end{align}  
where $x_{3/2}\equiv m_{\widetilde{G}}/m_{\widetilde{\chi}_1^0}$. The photon can also initiate the air shower, but as mentioned above, we will only consider the $\tau$ final state. For the parameter space we work with, the 3-body decay rate given by Eq.~\eqref{decaywidth2} is larger than the 2-body decay rate given by Eq.~\eqref{decaywidth} for very light gravitino with mass $m_{\widetilde{G}}\lesssim 0.1$ eV, which is actually preferred by cosmology~\cite{Dreiner:2011fp}. In particular, our scenario is naturally free from the cosmological gravitino problem~\cite{Moroi:1993mb} and consistent with cosmological constraints, such as from Lyman-$\alpha$ forest~\cite{Viel:2005qj}, cosmic microwave background (CMB) lensing and large-scale structure~\cite{Osato:2016ixc}.

%$\rm D\sim (2\rm R_{Earth}+h)\ \rm cos(\theta)-\rm h/ cos(\theta)\sim 6.575\times10^3$ km with uncertainty $\Delta \rm D\sim1.53\times10^3$ km (mainly due to the difference of incoming angle bewteen the two events detected\cite{Gorham:2018ydl}). From this geomotry requirement, we can get the decay time requirement straightforwardly as $\tau_{\chi}\sim0.219
%\pm 0.051
%$ ns. Combining this condition with Eq.~\eqref{decaywidth} we could easily get a geometry constraint for paramters $(\lambda_{\rm i31},\rm M_{\chi})$, which will be discussed further in next section.

\subsection{$LQD$ Contribution}
Now we consider the $\lambda'$-terms in Eq.~\eqref{eq:sup} which, when expanded, lead to the Lagrangian
\begin{align}
{\cal L}_{LQD} \ = \ & \lambda'_{ijk}\bigg[\widetilde{\nu}_{iL}\bar{d}_{kR}d_{jL}+\widetilde{d}_{jL}\bar{d}_{kR}\nu_{iL}+\widetilde{d}_{kR}^*\bar{\nu}_{iL}^c d_{jL} \nonumber \\
&  -\widetilde{e}_{iL}\bar{d}_{kR}u_{jL}-\widetilde{u}_{jL}\bar{d}_{kR}e_{iL}-\widetilde{d}_{kR}^*\bar{e}_{iL}^c u_{jL}\bigg]+{\rm H.c.}
\end{align}
These interactions will contribute to the neutrino-nucleon scattering mediated by either $s$ or $u$-channel exchange of a down-type squark, as shown in Fig.~\ref{fig:feynman} (bottom panel). For simplicity, we only consider the first-generation squark in the intermediate state. As for the initial state quarks, both $d$ and $s$ quark contributions turn out to be comparable. However, due to stringent constraints on the product $\lambda'_{i1k}\lambda'_{j2k}\lesssim 5\times 10^{-5}$ from $K$-meson  decays~\cite{Barbier:2004ez}, we will consider either the down-quark or the strange-quark in the initial state separately, but not both simultaneously. In particular, we will only consider the $\lambda'_{ijk}$ couplings with $j=1,2$ and $k=1$ for RH down-squark. After being resonantly produced, the bino will have a 3-body decay via off-shell down-type squark: $\widetilde{\chi}_1^0\to d \bar{d} \nu$ and $\widetilde{\chi}_1^0 \to u \bar{d} e$. %{\color{blue} (Jack: We need to edit the following. First and second generation will be equivalent in terms of ANITA XS, and second generation has weaker constraints. I advocate that we include both first and second generation constraints in fig 4 right, and here we should write we consider both generations.)} 
%\red{The final-state quarks readily hadronize, and for first-generation squark mediation.}
%\red{Since the down-quark parton distribution function (PDF) inside the proton is larger than the strange and bottom PDFs,} 
In this case, the final-state quarks from the 3-body bino decay  hadronize to either pions or kaons, mimicking the hadronic shower induced by the $\tau$. All other supersymmetric particles (except for the bino NLSP and gravitino LSP) are assumed to be heavier and not to play any role in our analysis. 

The total differential cross section for the (anti)neutrino-nucleon interactions can be written in terms of the Bjorken scaling variables $x=Q^2/2m_NE'_\nu$ and $y=E'_\nu/E_\nu$, where $m_N=(m_p+m_n)/2$ is the average mass of the proton and neutron for an isoscalar nucleon, $-Q^2$ is the invariant momentum transfer between the incident neutrino and outgoing bino, and $E'_\nu=E_\nu-E_{\widetilde{\chi}_1^0}$ is the energy loss in the laboratory frame. Keeping only the dominant $s$-channel contributions, we obtain~\cite{Carena:1998gd} 
 \begin{align}
%%%%%%%%%%%%%%%%%%%%%%neutrino
\frac{d\sigma^\nu_{LQD}}{dxdy} \ = \ &\frac{m_NE_{\nu}}{16\pi}\frac{|\lambda'_{ijk}|^2g'^2}{18}\left[ \frac{4xf_{d}(x,Q^2)}{\left(xs-m_{\tilde{d}_{R}}^2\right)^2+m^2_{\tilde{d}_{R}}\Gamma^2_{\tilde{d}_{R}}}\right. \nonumber \\
& \qquad + \left. \frac{xf_{\bar{d}}(x,Q^2)}{\left(xs-m_{\tilde{d}_{L}}^2\right)^2+m^2_{\tilde{d}_{L}}\Gamma^2_{\tilde{d}_{L}}}\right] \, , \label{lqd1} \\
%%%%%%%%%%%%%%%%%%%%%%%%%%%%%%%%%%anti neutrino
\frac{d\sigma^{\bar{\nu}}_{LQD}}{dxdy}\ = \ &\frac{m_NE_{\nu}}{16\pi}\frac{|\lambda'_{ijk}|^2g'^2}{18}\left[ \frac{xf_{d}(x,Q^2)}{\left(xs-m_{\tilde{d}_{L}}^2\right)^2+m^2_{\tilde{d}_{L}}\Gamma^2_{\tilde{d}_{L}}}\right. \nonumber \\
& \qquad + \left. \frac{4xf_{\bar{d}}(x,Q^2)}{\left(xs-m_{\tilde{d}_{R}}^2\right)^2+m^2_{\tilde{d}_{R}}\Gamma^2_{\tilde{d}_{R}}}\right] \, , 
\label{lqd2}
\end{align} 
where $s=2m_NE_\nu$ is the squared center-of-mass energy, and $f_d,f_{\bar{d}}$ are the PDFs for down and anti-down quark within the proton, respectively.  The Breit-Wigner resonance is regulated by the squark widths\footnote{The RH down-squark has two RPV decay modes: $\widetilde{d}_{kR}\to \nu_{iL}d_{jL}, \,  e_{iL}u_{jL}$, whereas the LH down-squark has only one: $\widetilde{d}_{kL}\to \nu_{iL}d_{jR}$. Similarly, for the $R$-parity conserving decays $\widetilde{d}\to d\widetilde{\chi}_1^0$, the RH squark coupling is twice that of the LH squark (due to different hypercharges).}
\begin{align}
\Gamma_{\tilde{d}_{kR}} \ = \ \frac{m_{\widetilde{d}_{kR}}}{8\pi}\left[\sum_{ij}|\lambda'_{ijk}|^2+\frac{2}{9}g'^2\right] \, , \\
\Gamma_{\tilde{d}_{kL}} \ = \ \frac{m_{\widetilde{d}_{kL}}}{16\pi}\left[\sum_{ij}|\lambda'_{ijk}|^2+\frac{1}{9}g'^2\right] \, .
\end{align}
Note that the resonance condition is satisfied for the incoming neutrino energy $E_\nu=m^2_{\widetilde{d}}/2m_Nx$, but due to the spread in the initial quark momentum fraction $x\in [0,1]$, the resonance peak is broadened and shifted above the threshold value $E_\nu^{\rm th}=m^2_{\widetilde{d}}/2m_N$, unlike the $LLE$ case where the resonance was much narrower. This is one of the reasons why the $LQD$ case allows for a larger parameter space than the $LLE$ case in explaining the ANITA events, as we show in the next section.   

%%%%%%%%%%%%%%%%%%%%%%%%%%%%%%%%%%%%%%%%%
\section{Event Rate}\label{s3}
%%%%%%%%%%%%%%%%%%%%%%%%%%%%%%%%%%%%%%%
We estimate the total number of expected events in the following way:  
\begin{equation}\label{eventNest}
N  \ = \ \int\limits_{E_i}^{E_f} d E_{\nu} \ \langle A_{\rm eff}\cdot\Delta\Omega \rangle \cdot T \cdot \Phi_{\nu} \, ,
\end{equation}
where $\Delta E \equiv E_f-E_i$ is the incident neutrino energy range that gives rise to the resonance, $\Phi_{\nu}(E_\nu)$ is the incoming neutrino flux, {$T = 53$ days is the total effective exposure time for the reported three flights of ANITA},\footnote{This corresponds to the combination of 17.25, 28.5 and 7 days of effective full-payload exposure time for ANITA-I, ANITA-II and ANITA-III, respectively, based on the experimental analysis given in Refs.~\cite{Schoorlemmer:2015afa,Gorham:2016zah,Gorham:2018ydl}. We have included the ANITA-II exposure time, even though it did not use a dedicated trigger algorithm sensitive to these events~\cite{Gorham:2016zah}.} 
and $\langle A_{\rm eff}\cdot\Delta\Omega \rangle$ is the effective area integrated over the relevant solid angle, averaged over the probability for interaction and decay to happen over the specified geometry shown in Fig.~\ref{fig:sketch}. 

For the $LLE$ case, since the resonance is very narrow, we can evaluate the energy integral $\int dE_\nu$ at the resonance energy $E_\nu=m^2_{\widetilde{\tau}}/2m_e$ with the energy spread $\Delta E=2\Gamma_{\widetilde{\tau}}m_{\widetilde{\tau}}/m_e\sim 0.05$ EeV.  The integrated effective area is defined as
\begin{align}
&\langle A_{\rm eff}\cdot\Delta\Omega \rangle \ \equiv \  \pi \theta_c^2 \int d\theta d\phi  \tan\theta \int_{\rm 0}^{l_{\rm tot}}\frac{dl_1}{l_{\rm BSM}}e^{-l_1\left(\frac{1}{l_{\rm BSM}}+\frac{1}{l_{\rm SM}}\right)} \nonumber \\
& \qquad \times 
\int_{l_{\rm tot}-D-l_1}^{l_{\rm tot}-l_1}\frac{dl_2}{l_{\rm decay}}e^{-\frac{l_2}{l_{\rm decay}}} \ (l_{\rm tot}-l_1-l_2)^2
\label{eq:effA}
\end{align}  
where $l_1$, $l_2$ are the distance traveled by $\nu$ and $\widetilde{\chi}_1^0$ respectively, and $l_{\rm tot}$ is the total distance from neutrino entering Earth to the detector, $\theta$ is the angle between the travel path and the vertical direction as defined before in Fig.~\ref{fig:sketch}, $D$ is the distance between the bino's decay point to the detector, $\theta_c\sim0.015$ is the Cherenkov cone angle, $l_{\rm SM}$ and $l_{\rm BSM}$ stand for the neutrino interaction lengths in the SM and BSM case, respectively. The interaction length can be generically written as $l_{\rm int}\sim 1/(\sigma N_A\rho)$, where $N_A$ is the Avogadro number, $\rho$ is the density and $\sigma$ is the interaction cross section. In our case, $l_{\rm BSM}$ is a function of the new physics parameters $\lambda_{ijk}$ and $m_{\widetilde{\tau}}$. Including the $\tau$-lepton decay probability would modify Eq.~\eqref{eq:effA} to the following: 
\begin{align}
&\langle {\rm A_{eff}\cdot\Delta\Omega} \rangle\equiv 
\pi \theta_{c}^2 
\int d\theta\ d\phi \tan\theta \int_{0}^{l_{\rm tot}}\frac{dl_1}{l_{\rm BSM}}e^{-l_1\left(\frac{1}{l_{\rm BSM}}+\frac{1}{l_{\rm SM}}\right)}\nonumber \\
&\qquad \times\int_{l_{\rm tot}-l_1-D}^{l_{\rm tot}-l_1} \frac{dl_2}{l_{\rm decay}}e^{-\frac{l_2}{l_{\rm decay}}}
\int_{0}^{D}\frac{dl_3}{l_{\rm decay,\tau}}e^{-\frac{l_3}{l_{\rm decay,\tau}}} \nonumber \\
& \qquad \times (l_{\rm tot}-l_1-l_2-l_3)^2
\end{align}
where $l_{\rm decay,\tau}$ is the decay length of $\tau$.  

\begin{figure}[t!]
	\centering
	\includegraphics[width=0.4\textwidth]{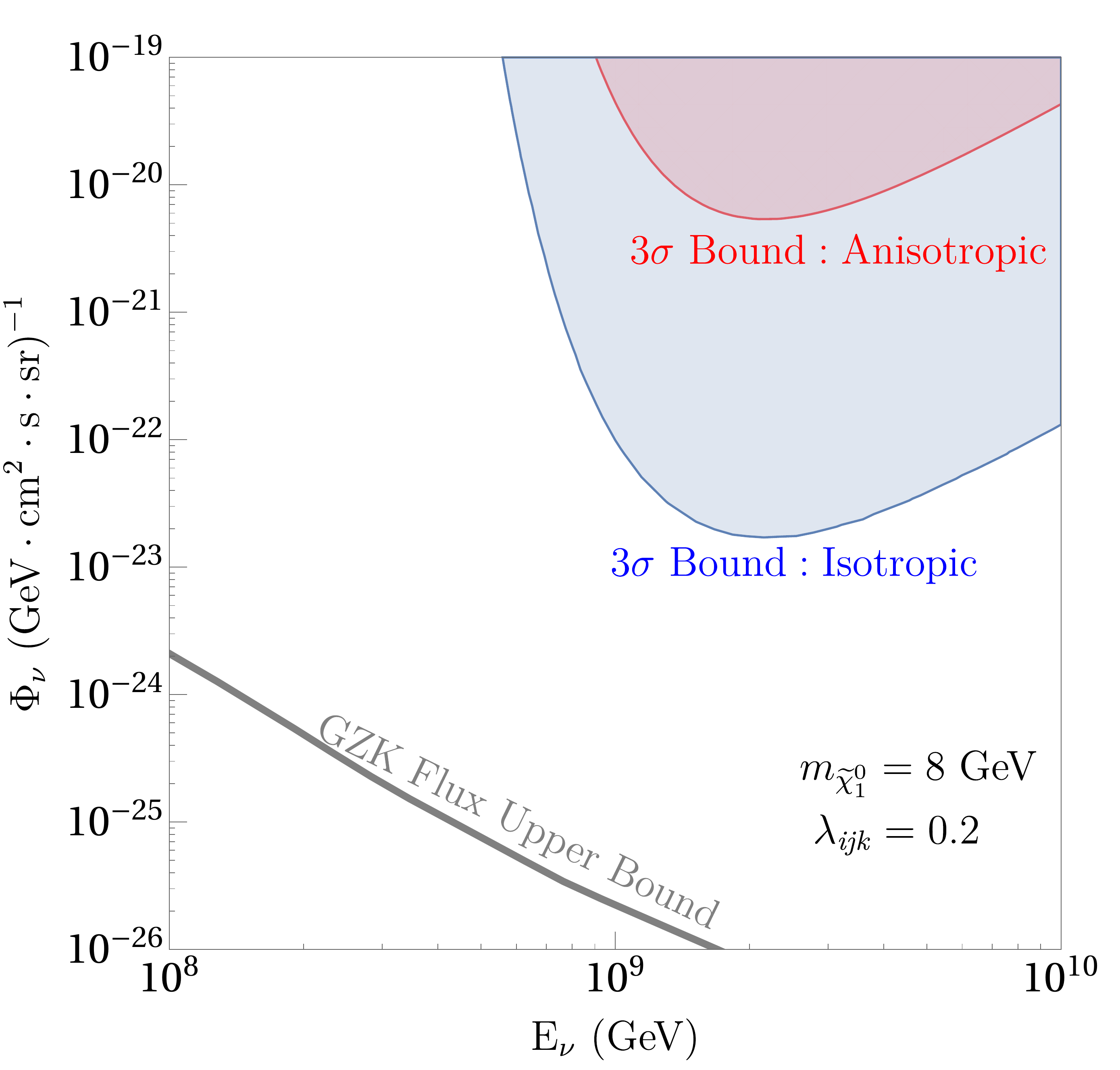}
	\caption{The minimum neutrino flux needed to explain the two ANITA events at $3\sigma$ confidence for an isotropic (blue shaded) and anisotropic (red shaded) area. The grey curve is the 90\% confidence level (CL) upper bound on the GZK flux from IceCube~\cite{Aartsen:2018vtx}. Here we have chosen $m_{\widetilde{\chi}_1^0}=8$ GeV and $\lambda_{ijk}=0.2$.}
	\label{fig:fluxconstriant}
\end{figure}

A quick test could be done to see if the GZK flux would be strong enough to provide the required number of events. Taking a benchmark point $m_{\widetilde{\chi}_1^0}=8$ GeV and $\lambda_{ijk}=0.2$, the flux needed to generate 2 events at 3$\sigma$ confidence (with Poisson distribution) is shown in Fig.~\ref{fig:fluxconstriant}.
The blue shaded region corresponds to the isotropic flux needed to match ANITA events within $3\sigma$. Thus, we need an isotropic flux as strong as $\sim5\times10^{-23}\ \rm (GeV\cdot cm^2\cdot s\cdot sr)^{-1}$, at least three orders of magnitude larger than the GZK upper limit $\sim 2.2\times10^{-26} \ \rm (GeV\cdot cm^2\cdot s\cdot sr)^{-1}$ at EeV level~\cite{Aartsen:2018vtx}, shown as the grey curve in Fig.~\ref{fig:fluxconstriant} . Therefore, under this RPV-SUSY bino scenario, GZK flux is disfavored as the source of UHE neutrinos at more than $3\sigma$ CL.

The challenge in explaining the ANITA event rate in terms of an isotropic GZK flux is quite general. Given the GZK upper limit of $\sim 50$ EeV beyond which the UHECR flux is suppressed due to interactions of UHECRs with relic photons~\cite{Greisen:1966jv, Zatsepin:1966jv} and noting that the average neutrino energy is roughly 5-10\% of the primary CR energy~\cite{Kelner:2006tc}, we can integrate the projected differential GZK flux given in Ref.~\cite{Kotera:2010yn} from 0.1~EeV up to 5~EeV, and use Eq.~\eqref{eventNest}, with $\Delta \Omega = 2 \pi$ and writing $A_\text{eff} = (4 \, \text{km}^2) \times \epsilon$ where $4 \, \text{km}^2$ is the inferred area of the radio cone for the observed ANITA events and $\epsilon$ is interpreted as a conversion efficiency for incoming neutrinos into observed upward going events at ANITA. We find a predicted event rate of $N \sim 600 \, \epsilon$. Two events at ANITA therefore suggests $\epsilon \sim 3 \times 10^{-3}$. Under the hypothesis that the event is initiated by a 
long-lived particle with $\gamma c \tau \sim R_\text{Earth}$ which decays below an altitude of 10~km after emerging from Antarctica (otherwise the air density drops rapidly and the shower does not fully develop before it reaches ANITA), there is already a suppression factor in $\epsilon$ which goes like $(10 \; \text{km}) / R_\text{Earth} \simeq 2 \times 10^{-3}$. Similarly, if the events are due to a decaying $\tau$ which has been produced in the collision between a highly energetic weakly interacting particle with scattering length $1/(\sigma N_\text{Av} \rho)\sim  R_\text{Earth}$ and a nucleus in the Antarctic crust or ice within 10~km of the surface, there is also the same suppression factor. This is before considering the production cross section for the long-lived particle in a $\nu\text{-}N$ collision in the northern hemisphere, and any additional branching ratio suppression.
% In fact, the test here is also valid for other possible new physics senario as long as GZK flux is assumed to be the source. For any possible new physics interactions, we could always have $\rm N\sim \pi\ \theta_{c}^2\  h^2/\rm cos[\theta]\cdot \Delta\Omega\cdot\Delta E\cdot T\cdot \Phi_{\nu}\sim 0.0064$ even if we take $\Delta \Omega=2\pi$, $\rm \Delta E=1$ EeV and assume all the neutrinos go through the new physics process in 100\% probability. Therefore, even under the most optimistic estimation, we still have the total event number outside the range of $3\sigma$ range ($\rm N>0.046$).

We can still consider anisotropic sources for the incoming neutrinos. In this case, the solid angle considered now ($\sim 0.0007\pi$, defined using the uncertainty of angles in the ANITA events~\cite{Gorham:2018ydl}) is much smaller than the solid angle in isotropic case ($\sim 1.3\pi$). Therefore, to get the same amount of events, the angular averaged anisotropic flux needed will become even larger than the isotropic flux, which is shown in Fig.~\ref{fig:fluxconstriant} as the red area. Due to the angular average effect, we can see that the required anisotropic flux is at least two orders of magnitude larger than the corresponding isotropic flux, i.e. $\sim 10^{-20} \ \rm (GeV\cdot cm^2\cdot s\cdot sr)^{-1}$. According to Refs.~\cite{Jacobsen:2015mga,Kalashev:2014vya,Mertsch:2016hcd,Aartsen:2016oji,Adrian-Martinez:2015ver}, such a large flux could in principle come from a transient point source, such as blazar, supernova burst, gamma-ray burst, or starburst galaxy, in the northern sky. In Refs.~\cite{Adrian-Martinez:2015ver, Aartsen:2016oji}, upper bounds on the strength of neutrino flux from point sources in the north sky are given as $\sim 3.2\times 10^{-20} \ \rm (GeV\cdot cm^2\cdot s\cdot sr)^{-1}$ around 0.5 EeV. %We also need to note that along the line of sight for the two events observed by ANITA, there might be more than one possible point source aligning in the same direction, making our current estimate a conservative one.
The ANITA collaboration~\cite{Gorham:2018ydl} also considered the transient source possibility for their anomalous events and, in fact, found a supernova candidate well within their expected angular uncertainty in the sky. The current and future ground arrays, such as Pierre Auger, Telescope Array (TA), AugerPrime and TA$\times$4 should be able to shed more light on these transient sources~\cite{Fujii:2018yhp}.   

\begin{figure*}[t!]
	\centering
	\includegraphics[width=0.49\textwidth]{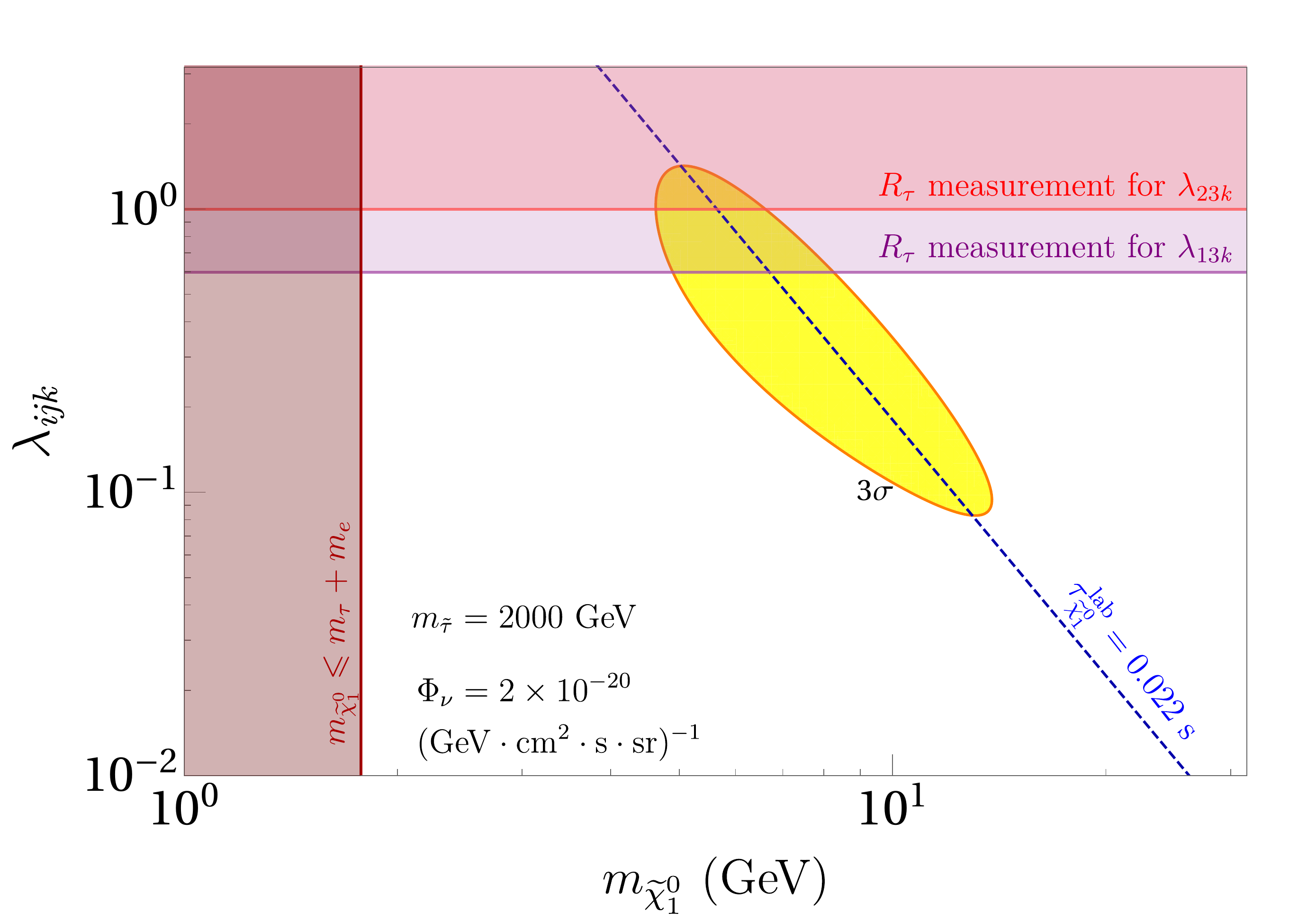}
\includegraphics[width=0.49\textwidth]{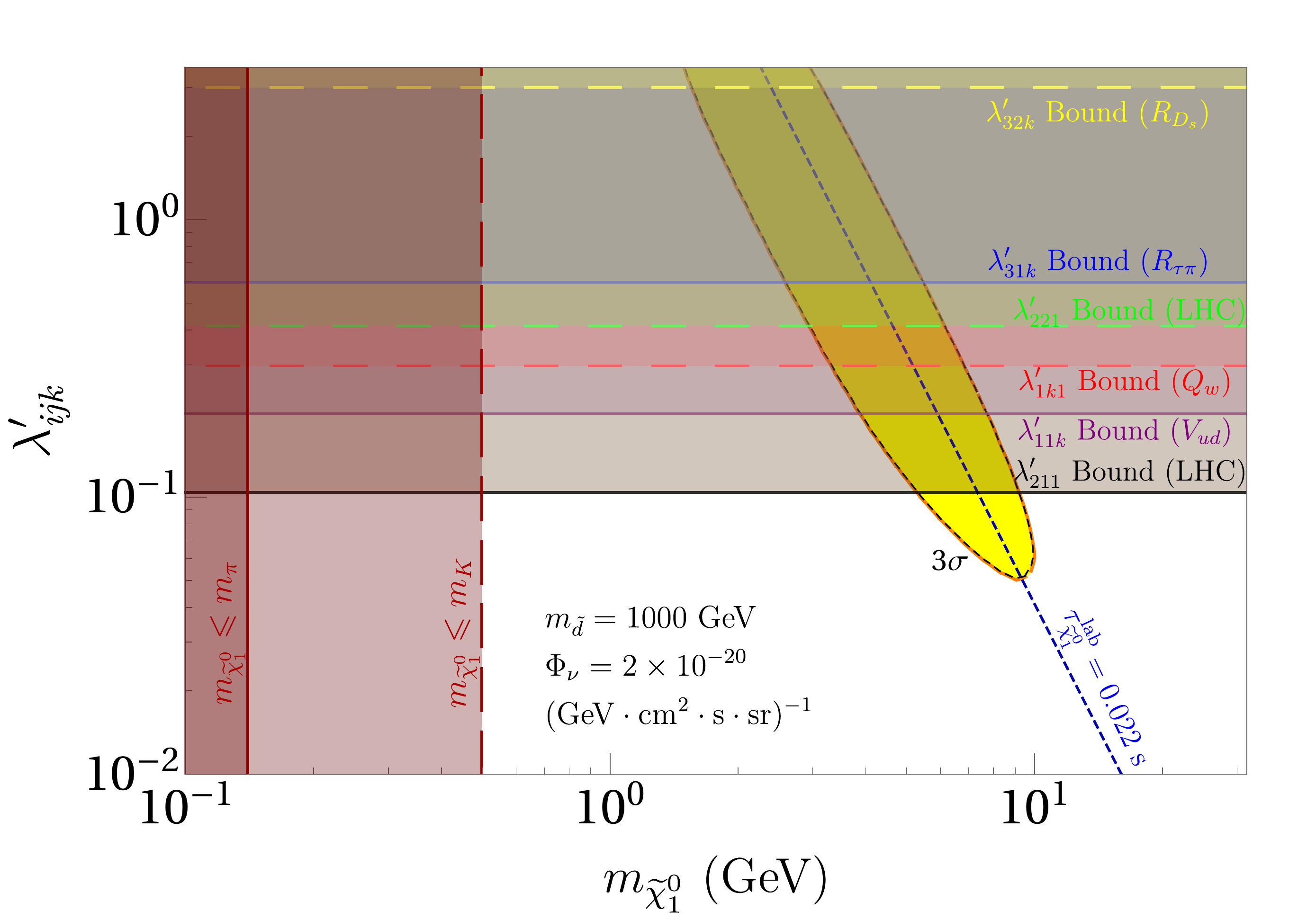}
	\caption{The 3$\sigma$ preferred region (yellow shaded) explaining the ANITA anomalous events in our RPV-SUSY framework for an average anisotropic neutrino flux of $2\times10^{-20}\  (\rm GeV\cdot cm^2\cdot s\cdot sr)^{-1}$. The left panel is for the $LLE$ case with a stau mass of $m_{\widetilde{\tau}}=2$ TeV, and the right panel is for the $LQD$ case with a down-squark mass of $m_{\widetilde{d}}=1$ TeV. In the right panel, the dashed contours are for $\nu-s$ initial state, while the solid contours are for $\nu-d$ initial state. The dashed blue line in each plot corresponds to the mean lifetime of the bino $\tau_{\widetilde{\chi}_1^0}^{\rm lab}=0.022$ s, obtained from setting the $\widetilde{\chi}_1^0$ decay length to match the average chord length. The horizontal shaded areas are the excluded regions for single RPV couplings from low-energy experiments~\cite{Kao:2009fg}. The vertical shaded regions are the kinematically forbidden regions for the bino decay considered here. }
	\label{fig:lambdaMconstriant}
\end{figure*}

Assuming the average strength of the anisotropic sources being $\Phi_\nu\sim 2\times10^{-20}\  (\rm GeV\cdot cm^2\cdot s\cdot sr)^{-1}$ with the mass of the RPV-SUSY mediator stau at $\rm m_{\widetilde{\tau}}=2$ TeV, we use Eqs.~\eqref{RPVXS} and \eqref{eventNest} and perform a statistical analysis to find the %$2\sigma$ and
$3\sigma$ favored region of the parameter space in the ($\lambda_{ijk}, m_{\widetilde{\chi}_1^0}$) plane, which is shown in the left panel of Fig.~\ref{fig:lambdaMconstriant} as the yellow shaded region. The dashed blue line shows the relation between the parameters once we set the bino decay length to be exactly the maximum distance traveled inside Earth, which corresponds to a mean lab-frame lifetime of $\tau_{\widetilde{\chi}_1^0}^{\rm lab}\sim 0.022 
%\pm 0.051
$ s. The horizontal purple and red shaded regions in Fig.~\ref{fig:lambdaMconstriant} are excluded from the $R_\tau$ measurements~\cite{Kao:2009fg}. The vertical shaded region is the kinematically forbidden region for the bino to decay into a $\tau$-lepton. Combining all the constraints, we find a window with $\lambda_{\rm i31}\sim 0.3$ and $m_{\widetilde{\chi}_1^0}\sim 8$ GeV for the new physics parameters to explain the events observed by ANITA. 
%Changing the stau mass to lower values would shrink the allowed parameter space, as it decreases the incoming neutrino energy range allowed by the resonance condition. We cannot increase the stau mass to higher values either, as that would require a higher energy for the incoming neutrino, which is already at the edge of the GZK limit.
The stau mass is roughly fixed to lie in the $1\text{--}2 \; \text{TeV}$ region by the requirement that $m_{\tilde{\tau}} = \sqrt{s} = \sqrt{2 m_e E_\nu}$, and that $E_\nu$ should be a few times larger than the observed cosmic ray energy of $0.2\text{--}1 \; \text{EeV}$. Such a particle may be observed in current or future collider experiments. The current LHC lower limits on the stau mass in the RPV-SUSY scenario is about 500 GeV, derived from multilepton searches~\cite{Aad:2014iza}. 

As for the $LQD$ case, we can do a similar calculation as for the $LLE$ case described above, except that we can no longer replace the energy integral in Eq.~\eqref{eventNest} by a delta function, due to a much broader resonance [cf.~Eqs.~\eqref{lqd1} and \eqref{lqd2}].  Our results for the $3\sigma$ preferred region are shown in the right panel of Fig.~\ref{fig:lambdaMconstriant} (yellow shaded region) for both $\nu-d$ (solid contours) and $\nu-s$ (overlapping dashed contours) initial states. The vertical shaded regions are the kinematically forbidden regions for the bino to decay into pions or kaons, corresponding to the $\lambda'_{i11}$ or $\lambda'_{i21}$ scenario, respectively. 
The horizontal shaded regions bounded by the purple and blue solid lines are excluded from the $V_{ud}$ and $R_{\tau\pi}$ measurements, respectively~\cite{Kao:2009fg}, which constrain the $\nu-d$ scenario. Similarly, the shaded regions bounded by the red and yellow dashed lines are excluded from the $Q_w$ and $R_{D_s}$ measurements, respectively~\cite{Kao:2009fg}, which constrain the $\nu-s$ scenario. Here we have chosen the RH down-squark mediator mass as $m_{\widetilde{d}}=1$ TeV. We do not include the LH squark contribution, because according to our estimates, the production cross section for a 1-TeV RH down-squark at the $\sqrt s=13$ TeV LHC is 6.2 fb, which is safely below the current upper limit of 13.5 fb~\cite{Aaboud:2017vwy}. Including the LH squark contribution increases the cross section to 15.5 fb. The black and green shaded regions in Fig.~\ref{fig:lambdaMconstriant} are the exclusion regions based on a recent update of the LHC constraints on the LQD couplings $\lambda'_{211}$ and on $\lambda'_{221}$, respectively~\cite{Bansal:2018dge}. 

Based on these bounds, we find that there is allowed parameter space at 
%both $2\sigma$ and 
$3\sigma$ for the $\lambda'_{i21}$ scenario ($\nu-s$ initial state), whereas for the 
$\lambda'_{i11}$ scenario ($\nu-d$ initial state), there is a smaller $3\sigma$ range with $\lambda'_{\rm i11}\sim 0.07-0.1$ and $m_{\widetilde{\chi}_1^0}\sim 7-10$ GeV  allowed. Increasing the squark mass moves the 
%$2\sigma$ and 
$3\sigma$ contours to larger $\lambda'$ values, which are excluded by the $V_{ud}$ and $Q_w$ measurements~\cite{Kao:2009fg}. Thus, we predict that if our $LQD$-type RPV-SUSY interpretation of the ANITA events is correct, then a TeV-scale squark should be soon found at the LHC.    Another independent test of the allowed parameter space shown in Fig.~\ref{fig:lambdaMconstriant} might come soon from the Belle II upgrade~\cite{Kou:2018nap}, which could significantly improve the $R_\tau$ measurements.  

Note that there is no LEP limit on our light bino scenario, because the $Z$ decay to binos is forbidden at the tree level. In fact, there is no lower limit for the bino mass, as long as it is not the dark matter candidate~\cite{Dreiner:2009ic}. Similarly, the stringent cosmological bounds on RPV couplings from the requirement that any baryon or lepton number violating interactions in equilibrium down to the electroweak scale could spoil the mechanism of baryogenesis due to fast electroweak sphaleron processes~\cite{Campbell:1990fa} can be avoided in our setup, because the mediator slepton/squark mass is at the TeV-scale and a low-scale baryogenesis could happen after they freeze out.

%Since this is a preliminary estimate based on a preselected average flux, the actual range for new physics coupling $\lambda$ might vary when we change the flux $F_\nu$. However, the allowed region of bino mass $\rm M_{\chi}$ is not so flexible($\precsim 10$ GeV) since it is constraint by the over lap of the 'Event Number Bound' and the 'Geometry Constriant'.

\section{IceCube Signatures}\label{icecube}
Explanations for the ANITA anomalous events which proceed via the decay of a $\tau$ lepton predict the presence of throughgoing track events at IceCube. While a few events exist which may be interpreted as being of this kind~\cite{Fox:2018syq,Kistler:2016ask}, their energies are one to two orders of magnitude smaller than those inferred for the events at ANITA. It is therefore worth exploring models which predict fewer or no throughgoing track events at IceCube. For both of the models presented in Section~\ref{sec:model}, only a fraction of events will proceed via a $\tau$ decay. A variation on the $LQD$ model may produce no ice-penetrating charged lepton signature at all. For example, a model making use of a $L_i Q_3 D_k$ vertex would mean no $\widetilde{\chi}_1^0 \to t \bar{d}_k \ell_i$ decay for $m_{\widetilde{\chi}_1^0} < m_t$, while a $L_1 Q_j D_k$ would lead to an electron which does not penetrate ice. In this case, the leading IceCube signature is $\widetilde{\chi}_1^0$ decay within the volume of the IceCube detector, which is suppressed in rate compared to ANITA by an additional factor of $h_\text{IC} / h_\text{ANITA} \sim 1/10$, in comparison with the throughgoing track signature. %However, should more events at ANITA be found, we would expect 

\section{Conclusion}\label{sec:conclusion}
We have explored a RPV-SUSY interpretation of the two anomalous upgoing air showers seen by ANITA. In our framework, the UHE neutrino interacts with Earth matter to resonantly produce a squark/slepton, which then decays to a long-lived bino, whose decay products are responsible for the upgoing air shower.  We considered both $LLE$ and $LQD$-type interactions and our main results are given in Fig.~\ref{fig:lambdaMconstriant}. We find that a light bino of a few GeV mass and the RPV couplings of order 0.1 provide the best-fit solution to the ANITA events. In the $LLE$ case, a stau of mass around 2 TeV, and in the $LQD$ case, a down-type squark of mass around 1 TeV are predicted, which should be accessible by the next run of the LHC. The Belle II upgrade will provide a complementary low-energy probe of the allowed parameter space. Our hypothesis could be completely tested with more events at ANITA-IV (and beyond), as well as by IceCube in the future. It would be remarkable if weak-scale supersymmetry was discovered in such an unexpected way! \\   

%In this article, we have explored the possibility of using RPV SUSY model to explain the two EeV-level upgoing ANITA events, which are unlikely to be explained in SM framework due to the contradiction between the long travel distance through Earth and the relatively short SM interaction length with Earth matter for EeV neutrinos. We proposed that a LSP bino $\chi$ is produced by $\nu\ e$ interaction through s-channel resonance with a NLSP $\tilde{\tau}$ as mediator. The bino is long-lived and relatively stable so that it could penetrate Earth before decaying back to leptons and get detected by ANITA. During testing, we find that GZK isotropic flux cannot provide enough neutrinos in either our RPV SUSY senario or any general new physics senarios that rely on isotropic astrophysical neutrino source. Even the most optimistic estimate giving $\rm N\sim 0.19$ is too small for ANITA observation within $3\sigma$ CL. Alternatively, we assume the anisotropic sources as our neutrino providers such as SBG, AGN and GC. With average flux set as $2\times10^{-20}\  (\rm GeV\cdot cm^2\cdot s\cdot sr)^{-1}$ and mediator mass at $\rm M_{\tilde{\tau}}=2$ TeV, we have found the allowed parameter space for $\lambda_{\rm i31}\sim0.15$ and $M_{\chi}\sim2$ GeV and shown them in Fig \ref{fig:lambdaMconstriant}. Within the allowed region, our model could provide explanation to ANITA events\cite{Gorham:2018ydl} as well as IceCube's null event\cite{Aartsen:2017mau} while being consistent with RPV coupling's current bound set by Barbar experiment\cite{Kao:2009fg}.

\acknowledgments
We gratefully acknowledge useful discussions with Dave Besson, Bob Binns, Slava Bugaev, JJ Cherry, Tony Gherghetta, Marty Israel, Kunio Kaneta, Doug McKay, Keith Olive and Brian Rauch, and collaboration in the early stages of this work with Kim Berghaus and Ofri Telem. JHC and BD acknowledge the local hospitality at Washington University and University of Maryland, respectively, where part of this work was done. The  work  of  JHC was  supported  in  part  by  NSF  Grant No. PHY-1620074 and the Maryland Center for Fundamental Physics. The work of BD and YS is supported by the US Department of Energy under Grant No.  DE-SC0017987.

%\section*{References}

\end{document}